\documentclass[aps,twocolumn,showpacs,preprintnumbers,amsmath,amssymb,prl,
superscriptaddress,floatfix]{revtex4}

\usepackage{ulem}
\usepackage{graphicx}

\begin{document}

\title{The fate of 1D spin-charge separation away from Fermi points}
\author{Thomas~L.~Schmidt}
\affiliation{Department of Physics, Yale University, 217 Prospect
Street, New Haven, Connecticut 06520, USA}
\author{Adilet~Imambekov}
\affiliation{Department of Physics and Astronomy, Rice
University, Houston, Texas 77005, USA}
\author{Leonid~I.~Glazman}
\affiliation{Department of Physics, Yale University, 217 Prospect
Street, New Haven, Connecticut 06520, USA}
\date{\today}

\newcommand{\tpsi}{\tilde{\psi}}
\newcommand{\tV}{\tilde{V}}
\newcommand{\hpsi}{\hat{\psi}}
\renewcommand{\d}{\hat{d}}
\newcommand{\trho}{\tilde{\rho}}
\newcommand{\hrho}{\hat{\rho}}
\newcommand{\alphabar}{{\bar{\alpha}}}
\newcommand{\expct}[1]{\left\langle #1 \right\rangle}
\newcommand{\expcts}[1]{\langle #1 \rangle}
\renewcommand{\mod}{\textrm{mod }}
\newcommand{\etal}{\textit{et al.}}

\newcommand{\red}[1]{\sout{\bf #1}}

\begin{abstract}
We consider the dynamic response functions of interacting one
dimensional spin-$1/2$ fermions at arbitrary momenta. We build a
nonperturbative zero-temperature theory of the threshold
singularities using mobile impurity Hamiltonians. The interaction
induced low-energy spin-charge separation and power-law threshold
singularities survive away from Fermi points. We express the
threshold exponents in terms of the spinon spectrum.
\end{abstract}


\maketitle

The low-energy excitations of interacting spin-$1/2$
fermions confined to one dimension (1D) is well represented by
two collective bosonic modes. These modes are the quantized waves
of spin and charge densities. 
Their spectra are linear; the
corresponding velocities, $v_s$ and $v_c$, differ from each
other. A microscopic consideration~\cite{giamarchi03,GNT} of the
repulsive interaction between spinful fermions leads to $v_c >
v_s$.

The spectra of the collective modes can be probed in a
momentum-resolved tunneling~\cite{auslaender,jompol09} or in an
ARPES~\cite{kim06} or photoemission~\cite{photo_Jin} experiment.
In these  methods, a spin-$1/2$ fermion with a given momentum
tunnels into or out of the studied system. The tunneling inevitably perturbs each of the
two collective modes. For example, in the case of a low-energy particle ($k_F
\gg
k-k_F>0$), the small difference $k-k_F$ is shared between the
excitations of the two modes. The Luttinger liquid (LL) theory
predicts~\cite{GNT} that at given $k$ the tunneling probability
is singular at energies $v_s(k-k_F)$ and $v_c(k-k_F),$
corresponding to the entire momentum $k-k_F$ given to the {\it
``spinon''} or {\it ``holon''} belonging to the spin and charge
mode, respectively. The exponents of the two power-law
singularities depend on a single number, the LL parameter $K_c$
for the charge mode. The two sharp peaks in the momentum-resolved
tunneling probability at energies associated with excitation of
the two modes, $\omega=v_{s,c}\cdot (k-k_F),$ are the hallmark of
the spin-charge separation in the LL.

The momentum-resolved tunneling rate is proportional to
the fermionic spectral function  $A(k,\omega) = \tfrac{1}{\pi}
\text{Re} \int dt dx e^{i \omega t - i k x} \theta(t) \expct{ \{
\psi(x,t), \psi^\dag(0,0) \} }$. Recently, considerable progress was achieved in the analytical theory of
dynamic responses of a 1D system away from the Fermi points for
{\it spinless} fermions~\cite{pustilnik06,khodas07_2,
pereira09,imambekov09_2,imambekov09}. The developed methods map
the 1D dynamic response problem near the edge of support onto the
``mobile quantum impurity'' effective
Hamiltonian~\cite{balents00}. For spinless fermions in the weak-interaction limit, one may consider the generic spectrum of free
fermions exactly, while treating their interaction
perturbatively~\cite{pustilnik06,khodas07_2}. For example, at
$|k|<k_F$ the threshold coincides with the spectrum of a hole,
which can be thought of as a mobile quantum impurity. Because of the
interactions,  it ``shakes up'' the fermions in the vicinity of
Fermi points, leading to the orthogonality catastrophe and to the
power-law behavior of $A(k,\omega)$ at the
threshold~\cite{pustilnik06,khodas07_2}. The perturbation
theory allows one to identify the quantum numbers of the impurity
and to match the phenomenological theory of threshold
exponents~\cite{imambekov09} valid at any interaction strength with the
weak-interaction limit.

In this Letter we build a nonperturbative zero-temperature theory
of threshold singularities of dynamic responses of 1D spin-$1/2$
fermions at arbitrary $k,$ shedding light on the fate of the
spin-charge separation away from Fermi points. The obvious
difficulty of the problem lies in the appearance of distinctly
different spin and charge modes at any interaction strength.
For weak interactions, these modes are degenerate, which
renders perturbation theory inapplicable. We find that
interaction induced spin-charge separation survives away from the
Fermi points, and manifests itself in the choice of quantum
numbers of the mobile impurity: for repulsive interactions, it
only carries spin but no charge. We fix the parameters of the
effective quantum impurity Hamiltonian using $SU(2)$ and Galilean
invariance, and find the exponents of threshold singularities of
various dynamic responses [density and spin structure
factors as well as $A(k,\omega)$].

\begin{figure}[t]
  \centering
  \includegraphics[width = 0.48 \textwidth]{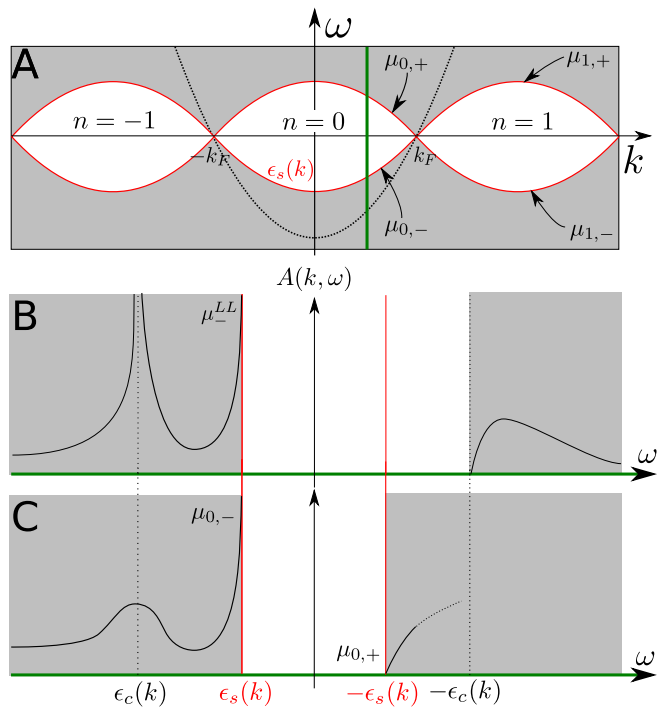}
\caption{(Color online) Structure of the spectral function
$A(k,\omega)$ in the $(k,\omega)$-plane (A) and along a cut for
fixed  $0<k < k_F$ (B,C). (A) Shaded areas indicate the regions
where $A(k,\omega)$ is nonzero. (B) Luttinger liquid (LL) results:
in the vicinity of the spinon mass shell, $A(k,\omega) \propto
(\omega - v_s k)^{-\mu_-^{LL}}$ where $\mu_{-}^{LL} =\tfrac{1}{2}-
\tfrac{1}{4} \left( K_c + K_c^{-1} - 2\right).$ (C) Schematic
behavior of the spectral function beyond the LL approximation
away from the Fermi points: the exponent $\mu_{0,-}$ is different from
$\mu_{-}^{LL}$, the charge mode is smeared out, and for $\omega>0$
the region of finite support starts from $-\epsilon_s(k)$ with an
exponent $\mu_{0,+}$ instead of from $-\epsilon_c(k).$ }
  \label{fig:a}
\end{figure}

Away from $k=k_F,$ the spectrum of the spinon mode $\epsilon_s(k)$
departs from the linear one, $v_s(k - k_F),$ and becomes a
periodic function of $k$ with period $2k_F.$ The threshold for
$A(k,\omega)$ is located at $|\omega_k|=\epsilon_s(k-2nk_F)$,
with integer $n$ such that $|k-2nk_F|\leq k_F,$ see
Fig.~\ref{fig:a}. The spin-charge separation is preserved near the
threshold at arbitrary $k$ in the following sense: if the energy
$\omega$ of, say, the extracted fermion approaches the threshold
$\omega_k$, then the momentum of a created spinon is approaching
$k-k_F.$ The rest of the energy, $\sim|\omega-\omega_k|\rightarrow
0,$ is given to a holon; it inevitably resides near a Fermi point
and may be described as a conventional linear LL. We express the
exponents of various threshold singularities in terms of the
derivatives of $\epsilon_s(k)$ with respect to $k$ and $\rho,$
the density of the liquid [see Eq.~(\ref{phen_final}) and
Table~\ref{tab:Exponents}]. The obtained exponents are valid at
arbitrary $k,$ including the Fermi points.

Near the Fermi points, the exponents for $A(k,\omega)$
approach the universal values which depend only on $K_c.$ In the
main region ($n=0$), the exponent for $A(k,\omega<0)$ coincides
with the predictions of the linear LL (unlike in the spinless
case~\cite{imambekov09_2}), while density and spin structure
factor exponents approach $1/2.$

Let us first introduce the
refermionization of the linear LL Hamiltonian. Its charge ($c$)
and spin ($s$) parts separate from each other, $H_0 = H_c + H_s$,
and in terms of boson variables, these parts have the
conventional form~\cite{giamarchi03},
\begin{align}\label{H_nu}
  H_\nu &= \frac{v_\nu}{2 \pi} \int dx \left[ K_\nu (\partial_x
    \theta_\nu)^2 + \frac{1}{K_\nu} (\partial_x \phi_\nu)^2 \right]
\end{align}
for $\nu = c,s$. The canonically conjugate fields $\partial_x
\theta_\nu$ and $\phi_\nu$ are the momentum and displacement
operators of bosonic charge and spin density waves and satisfy
$[\phi_\nu(x),\partial_y \theta_\mu(y)] = i \pi \delta_{\mu\nu} \delta(x-y)$.
 The effects of the interaction are
contained in the constants $K_\nu$, but most importantly
interaction yields a difference between the velocities of charge
and spin modes, thus removing the degeneracy characteristic of
the free-fermion system. For repulsive interactions at the
$SU(2)$ symmetric point, we can assume rather generally $K_s =
1$~\cite{giamarchi03,GNT}. Each of the Hamiltonians ($c,s$) may be
refermionized in terms of free fermionic quasiparticles, the same
way as it was suggested earlier~\cite{mattis65,rozhkov05} for
spinless particles. One obtains a free linear
Hamiltonian in terms of left- and right-moving fermionic spin and
charge quasiparticles $\tpsi_{\alpha\nu}(x)$ $(\alpha =
R,L=\pm)$, spinons and holons, respectively~\cite{rozhkov05}:
\begin{align}\label{MI_0}
  H_0 &= -i  \sum_{\nu = c,s} v_\nu \sum_{\alpha = R,L} \alpha \int dx :
\tpsi^\dag_{\alpha \nu}(x) \nabla
  \tpsi_{\alpha\nu}(x):.
\end{align}
The operators $\tpsi_{\alpha\nu}(x)$ have fermionic commutation
relations, and their vacuum state has unity occupation for
negative (positive) momenta for $\alpha=R(L).$
The left- and right-moving components of the original fermions,
defined by $\psi_\sigma(x) = e^{-i k_F x} \psi_{L\sigma} + e^{i
k_F x} \psi_{R\sigma}$, can be expressed in terms of refermionized
quasiparticle operators as $(\sigma = \uparrow,\downarrow = \pm)$
\begin{align}\label{psi_tpsi}
 \psi_{\alpha\sigma}(x)
&\propto
 \tpsi_{\alpha c}(x) F_{\alpha c}(x) \tpsi^{-\sigma}_{\alpha s}(x)
F^\sigma_{\alpha
 s}(x).
\end{align}
Here notations are such that $\tpsi^{\pm}_{\alpha s}(x)$ is the
creation (annihilation) operator of a spinon, and the string
operator $F_{\alpha s}(x)$ has to be raised to power $\sigma.$
We did not write out the Klein factors explicitly, since
they do not affect expectation values which define response
functions. For example, the correlation function $\expct{\{ \psi(x,t), \psi^\dag(0,0)\}}$
which defines $A(k,\omega)$ conserves the spinon and holon
numbers. The string operators
\begin{align}
  F_{\alpha\nu}(x) = \exp\left\{ -i \alpha \int_{-\infty}^x dy\ \left[
      \delta_{+\nu} \trho_{\alpha \nu}(y) + \delta_{-\nu}
      \trho_{-\alpha\nu}(y) \right] \right\} \notag
\end{align}
are functions of
the quasiparticle densities $\trho_{\alpha\nu}(x) =
:~\tpsi^\dag_{\alpha\nu}(x) \tpsi_{\alpha\nu}(x):$.
The effects of the interaction are contained in the phase shifts
$\delta_{\pm\nu}$, which are given by
\begin{align}
 \frac{\delta_{\pm\nu}}{2\pi} &= \left(\frac{1}{2} \pm \frac{1}{2}\right) \mp
\sqrt{\frac{1}{8 K_\nu}} - \sqrt{\frac{K_\nu}{8}}. \label{phases}
\end{align}
$K_s=1$ leads to $\delta_{-s} = 0$ and $\delta_{+s}/(2\pi) = 1 -
1/\sqrt{2}$. Within the linear spectrum approximation of the LL
theory, the dynamics of the string operators is linear, and
Eqs.~(\ref{MI_0})-(\ref{phases}) lead to the conventional LL
results.

We first discuss the deviations from the LL results for the
exponents of the spectral function at the charge mode in the
vicinity of the Fermi point $+k_F$. Similar to the spinless
case~\cite{imambekov09_2}, one can describe the exponents by taking
into account only the quadratic nonlinearity in the spectrum of
charge quasiparticles. Interactions between charge quasiparticles
lead only to small corrections to the exponent at the charge mode
for $|k-k_F|\ll k_F,$ and we obtain a singularity at the holon mass shell $A(k,\omega) \propto [\omega -
\epsilon_c(k)]^{-\mu_c},$ where
\begin{align}
 \mu_c = 1 - \left(\frac{\delta_{-c}}{2\pi}\right)^2 -
 \left(\frac{\delta_{+s}}{2\pi}-1\right)^2 -
 \left(\frac{\delta_{+c}}{2\pi}\right)^2.
\end{align}
The nonlinearity in $\epsilon_c(k)$ changes the exponents
compared to LL theory in an energy window of width $\sim
(k-k_F)^2$ around $\epsilon_c(k)$. Even at $K_c\rightarrow 1,$
the exponent $\mu_c \to \sqrt{2}-1,$ different from the LL
prediction $1/2$. Beyond this energy window, the usual LL
behavior is recovered.

On the other hand, interactions between spinons on the same branch
are important in the description of the response functions at the
spinon spectrum even for $|k-k_F|\ll k_F.$ The difference stems
from the fact that the leading-order band curvature of the
spinons is cubic \cite{footnote2} as the $SU(2)$ symmetry
enforces particle-hole symmetry for the spinon mode. If one
attempts now to treat the leading interactions between spinons
perturbatively in the spirit of Ref.~\cite{khodas07_2}, one finds
that these interactions lead to $\sim O(1)$ changes of the
exponents in the response functions, since the difference in the
velocities of spinons is proportional to $\propto (k-k_F)^2$
unlike $\propto (k-k_F)$ for holons.

Nevertheless, it is still possible to describe
the spectral function at the spinon mode using mobile
impurity models, as has been established in a number of articles~\cite{khodas07_2,imambekov09,imambekov09_2,pereira09,balents00}.
We shall first apply the approach to calculate $A(k,\omega)$ for
$|k-k_F| \ll  k_F, \omega<0$ in the vicinity of the spinon mode
$\omega \approx \epsilon_{s}(k)$. According to
Eq.~(\ref{psi_tpsi}), for energies close to $\epsilon_s(k)$, the
configurations of lowest energy will contain a spinon hole at
momentum $k_d = k - k_F<0$ and a  holon  at $k_F.$ Therefore, we
can project the Hamiltonian onto a band structure which consists
of a ``deep'' right-moving spinon hole $\d$ at momentum $k_d$ as
well as spinon and holon states, $\hpsi_s$ and $\hpsi_c$,
respectively, at the Fermi points. This leads to the Hamiltonian
$H_0+H_d+H_{int},$ where $H_0$ is given by Eq.~(\ref{MI_0}),
while the other terms read
\begin{align}\label{MI_0d}
  H_d &= \int dx\ \d^\dag(x) [ \epsilon_{s}(k) - i \hbar v_d \nabla ]
  \d(x),\notag \\
H_{int}& = \int dx \sum_{\alpha \nu} \tV_{\alpha\nu}(k)
\hrho_{\alpha \nu}(x) \d(x) \d^\dag(x).
\end{align}
The subbands were linearized around $k_d$ and the Fermi
points, and we used $\tpsi_{Rs} \propto \hpsi_{Rs} + e^{i k_d x}
\d$ and $\tpsi_{\alpha\nu} \propto \hpsi_{\alpha\nu}$ in all
other cases. The velocity of the hole $\d$ is given by $v_d =
\partial \epsilon_{s}(k)/\partial k$ and
$\tV_{\alpha\nu}(k)$ reflect the interaction
of the hole with the modes near the Fermi points \cite{footnote3}.

The interaction term (\ref{MI_0d}) can be removed using a unitary
transformation~\cite{khodas07_2,imambekov09,imambekov09_2,pereira09,balents00}.
This will lead to additional phase shifts $\Delta
\delta_{\alpha\nu} = \tV_{\alpha\nu}/(v_d - \alpha v_\nu)$ which
determine the edge exponents. Except for $\Delta \delta_{+s},$
these additional phase shifts are small since the corresponding
$\tV_{\alpha\nu}$ vanish for $k\rightarrow k_F,$ while the
denominator remains finite. The remaining phase shift $\Delta
\delta_{+s}$ can be fixed using the $SU(2)$
symmetry~\cite{imambekov09}. In particular, this symmetry
requires identical exponents of the spin correlation functions $
S^{-+}(k,\omega)$ and   $S^{zz}(k,\omega),$ where \textit{e.g.} $
  S^{zz}(k,\omega) = \int dt dx\ e^{i \omega t - i k x} \expcts{S^z(x,t)
  S^z(0,0)}.
$
We can calculate the exponents of these functions at the spinon
mass shell for general phase shifts $\delta_{+s}^* = \delta_{+s} +
\Delta \delta_{+s}$. The leading exponents for $
S^{-+}(k,\omega)$ and   $S^{zz}(k,\omega)$ read $ 1 -  \left[
1/\sqrt{2} \mp \delta_{+s}^*/(2\pi) \right]^2$, respectively.
Hence, the $SU(2)$ symmetry leads to $\delta_{+s}^* = 0$ and
rules out the interaction of the impurity with the
low-energy spinons, and the LL result $\mu_{-}^{LL}$(see
Fig.~\ref{fig:a}) for the exponent of the spectral function remains valid for $|k-k_F|\ll k_F.$

\begin{table*}
  \centering
  \begin{tabular}{|c|c|c|l|}
   \hline
   $A(k,\omega \gtrless 0)$ & $(2n-1) k_F < k < (2n + 1) k_F$ & $\mu_{n,\pm}$
      & $\displaystyle 1 - \frac{1}{2} \left( -\frac{(2n + 1)
\sqrt{K_c}}{\sqrt{2}} + \frac{\delta^A_+ + \delta^A_-}{2\pi} \right)^2 -
\frac{1}{2} \left( \frac{1}{\sqrt{2 K_c}} - \frac{\delta^A_+ - \delta^A_-}{2\pi}
\right)^2 - m^2_\pm$ \\
   \hline
   $S(k,\omega)$ & $2n k_F < k < 2(n + 1) k_F$ & $\mu_{n}^{DSF}$
          & $\displaystyle \frac{1}{2} - \frac{1}{2} \left( \frac{2n
\sqrt{K_c}}{\sqrt{2}} + \frac{\delta^S_+ + \delta^S_-}{2\pi} \right)^2 -
\frac{1}{2} \left(\frac{\delta^S_+ - \delta^S_-}{2\pi} \right)^2$ \\
   \hline
  \end{tabular}
  \caption{Exponents for the
   spectral function $A(k,\omega)$ (see Fig.~\ref{fig:a} for notations)
   and the dynamic density structure factor $S(k,\omega)$  at the edge of
support.
   The  exponents are determined in terms of the phase shifts
   $\delta^A_\pm = \Delta \delta_{\pm c}(k - 2 n k_F)$ and
   $\delta^S_\pm = \Delta \delta_{\pm c}[(2n + 1) k_F - k]$
   calculated in Eq.~(\ref{phen_final}). Moreover, $m_{\pm} = (n+1/2 \pm 1/2)\,
\mod 2.$
   The exponents for the spin structure factor $S^{zz}(k,\omega)$
   coincide with the exponents for $S(k,\omega)$.}
  \label{tab:Exponents}
\end{table*}

Even beyond the universal regime $|k-k_F| \ll k_F,$  the
exponents of the spectral function near its edge of support
survive, and  may be determined using mobile impurity
Hamiltonians. Above analysis in the limit $|k-k_F| \ll k_F$
indicates that in the vicinity of the Fermi points the impurity
has the quantum numbers of a spinon, and the impurity Hamiltonian
is given by Eq.~(\ref{MI_0d}). The continuous evolution of edge
exponents implies that the same Hamiltonian should describe the
edge exponents even significantly away from Fermi points. As a
consequence of such a spin-charge separation beyond low energies,
for Galilean-invariant systems we can  express all exponents in
terms of the edge position only, as in the spinless case
~\cite{imambekov09}. For this purpose, we rewrite Eq.~(\ref{MI_0d}) as
\begin{align}\label{Hint}
 H_{int} = \int dx \left[ V_R \nabla \frac{\theta_c - \phi_c}{2 \pi} - V_L
\nabla \frac{\theta_c + \phi_c}{2\pi} \right] \d(x)
 \d^\dag(x),\notag
\end{align}
since we saw  previously that the coupling constant to the spin
sector vanishes. The argument was based on $SU(2)$-symmetry and
is valid beyond the low-energy regime.

The interaction term is removed~\cite{imambekov09} by a unitary transformation $U^\dag (H_{c}+H_d+H_{int}) U$, where
$U= \exp \{ i \int dx [\frac{\Delta\delta_{+c}}{2\pi}(\theta_{c}\sqrt{K_c}-\frac{\varphi_c}{\sqrt{K_c}})-\frac{\Delta\delta_{-c}}{2\pi}(\frac{\varphi_c}{\sqrt{K_c}}+ \theta_{c}\sqrt{K_c})] \hat{d}(x) \hat{d}^{\dagger}(x) \}$ with phases $\Delta \delta_{\pm c}$ defined by
\begin{align}
 (V_L \mp V_R)K^{\mp 1/2}_c &= - \Delta\delta_{-c} (v_d + v_c) \pm \Delta\delta_{+c} (v_d - v_c). \notag
\end{align}
The fermionic operator can be written as $\psi_{\uparrow} = \hat d e^{i(\theta_c-\varphi_c)/\sqrt{2}}$ (with the second factor coming from a holon at $k\to k_F$), and then its correlations can be evaluated using the operator $U$ similar to the spinless case. The phases $\Delta \delta_{\pm c}$ fix the exponents of the spectral function as well as of the dynamic density and spin structure factors \cite{khodas07_2,imambekov09,imambekov09_2,pereira09,balents00}.

In order to relate $V_{L,R}$ to
the spectrum $\epsilon_s(k),$ we first
consider the shift in energy due to a change in density. A
uniform variation of the density by  $\delta \rho$ corresponds to
a finite expectation value $\expct{\nabla \phi_c} = - \pi \delta
\rho/\sqrt{2},$ and leads to the shift in the single-particle
energy
\begin{equation}\label{delta_epsilon_s1}
 \delta \epsilon_s(k) =  \left[\frac{\partial \epsilon_s(k)}{\partial \rho} +
\frac{\partial \mu}{\partial \rho} \right] \delta
 \rho,
\end{equation}
where $\mu$ denotes the chemical potential.  We now need to
calculate the same change in energy using Eq.~(\ref{MI_0d}). Unlike the spinless case~\cite{imambekov09}, the
spinon momentum changes under variation of the density, since the
total momentum is fixed while the holon momentum changes following
the shift of the Fermi point. Calculating the shift
in the energy of the spinon $\d$ as well as the shift in energy
of the holon at the Fermi
 point,  and comparing with Eq.~(\ref{delta_epsilon_s1}), one finds
\begin{align}\label{Vsum}
 -\frac{V_R + V_L}{2\sqrt{2}} = \frac{\partial \epsilon_s(k)}{\partial \rho} +
\frac{\pi}{2} \frac{\partial \epsilon_s(k)}{\partial
 k}.
\end{align}
The second relation can be derived using  Galilean invariance, which
predicts~\cite{imambekov09} that a uniform change in velocity $u$
should lead to a change in energy of
\begin{align}\label{delta_epsilon_s2}
 \delta \epsilon_s(k) = m u \left[ k - \frac{\partial \epsilon_s(k)}{\partial k}
 \right],
\end{align}
where $m$ is the bare mass. On the other  hand, this change in
velocity $u$ will lead to a finite expectation value,
$\expcts{\nabla \theta_c} = \sqrt{2} m u.$ Because of the shift of
the  Fermi point, this leads to an energy shift of a holon at
the right Fermi point $K_c v_c mu= v_F m u,$ and a spinon energy shift
$-muv_d$ due to the change of the spinon momentum. In addition, the
interaction Hamiltonian yields a shift
$\sqrt{2} m u (V_L - V_R)/(2 \pi)$. Combining these terms leads to
\begin{align}\label{Vdiff}
 \frac{V_L - V_R}{\sqrt{2}\pi} = \frac{k - k_F}{m}.
\end{align}
Equations (\ref{Vsum}) and (\ref{Vdiff})  allow us to express
$V_{L,R}$ in terms of the derivatives of the single-particle
spectrum $\epsilon_s(k)$. For the phase shifts, we thus obtain
the result
\begin{align}\label{phen_final}
 \frac{\Delta \delta_{\pm c}(k)}{2\pi} = \pm \frac{\frac{k - k_F}{m \sqrt{K_c}}
\pm \sqrt{K_c} \left( \frac{2}{\pi} \frac{\partial \epsilon_s(k)}{\partial \rho}
+ \frac{\partial \epsilon_s(k)}{\partial k} \right)}{2 \sqrt{2} \left(
\frac{2}{\pi} \frac{\partial \epsilon_s(k)}{\partial k} \mp \frac{k_F}{m K_c}
 \right)}.
\end{align}
The predictions (\ref{phen_final}) and the absence of
the coupling of the spinon impurity to the low-energy spinons
can be explicitly checked for the case of the integrable
Yang-Gaudin model based on its finite size
spectrum~\cite{essler09}. Moreover, for $k \to k_F$,
Eqs.~(\ref{Vsum}) and (\ref{Vdiff}) yield $V_L = V_R = 0$ which
reflects the absence of interactions between holons and spinons
within the linear LL theory.

Finally, let us present the exponents of the spectral function in the
regions $(2n - 1) k_F < k < (2 n + 1) k_F$ for integer $n.$ For
$\omega < 0$, in the vicinity of $\epsilon_s(k)$, the symmetry of
the edge position in $k$ leads to a spinon hole at momentum
$k_{d,n} = k - (2 n + 1) k_F < 0$, a  holon  at $k_F$ as well as
additional excitations which absorb the remaining momentum $2 n
k_F$. In terms of the original fermions, these ``umklapp''
excitations contain particles or holes
at the Fermi points, and are thus characterized by four
parameters. Fixing the total momentum at $2 n k_F$ and requiring
zero total charge and spin leaves one parameter $m_-,$ where the
umklapp contribution to  $\psi_{\sigma}$  can be represented as
$(\psi^{\dagger}_{R\uparrow}\psi_{L\uparrow})^{-
(n + m_-)/{2}}
(\psi^{\dagger}_{L\downarrow}\psi_{R\downarrow})^{(n
- m_-)/{2}},$
and thus $m_-$ has  to satisfy the selection rule $m_- \equiv n
\,(\mod 2)$. The exponent of the spectral function is now determined by
the phase shifts (\ref{phen_final}) taken at momentum $k - 2 n
k_F.$\cite{imambekov09} For $\omega > 0$, one can also describe the exponents using
the impurity Hamiltonian with the same parameters, but the states
which determine the exponents are different. Following the same
line of arguments as previously, one finds exponents which are
formally identical to the $\omega < 0$ case, but where $m_+$ now
has to satisfy a different selection rule, $m_+ \equiv (n+1)\,
(\mod 2)$. The leading exponents stem from $m_{\pm}$ with smallest
absolute value allowed by the selection rules, and the results
are shown in Table~\ref{tab:Exponents}.

The exponents of the spin structure factor $S^{zz}(k,\omega)$ and
the dynamic structure factor $S(k,\omega) = \int dt dx\ e^{i(\omega t
- k x)} \expct{\rho(x,t) \rho(0,0)}$  can also  be calculated
similarly. It turns out that the exponents for $S(k,\omega)$
coincide with those for $S^{zz}(k,\omega),$ and they are shown in
Table~\ref{tab:Exponents}.

In conclusion,  we have calculated the exponents of  dynamical
correlation functions of 1D spinful interacting fermionic systems
beyond the approximation of a linear spectrum, shedding light on
the fate of the spin-charge separation away from Fermi points. In
the low-energy sector near the Fermi points, we found universal phase
shifts which control the exponents and depend only on the LL
parameter $K_c$. Beyond the low-energy regime, we were able to
establish phenomenological relations between  phase shifts and
properties of the spinon spectrum.

\acknowledgments We thank F.~Essler and R.G. Pereira for useful
discussions. We acknowledge support by
the NSF DMR Grant No. 0906498, the Nanosciences Foundation at Grenoble, France, and the Swiss NSF.

{\it Note Added. -- } Recently, the preprint \cite{PereiraSela} appeared, where the exponents
for the density and spin structure factors for $k\ll k_F$ were derived.


\begin{thebibliography}{19}
\bibitem{footnote2} In general, the curvature is negative.
\bibitem{footnote3} We checked that marginally irrelevant \cite{giamarchi03}
spin-flip scattering does not affect the parameters in Eq.~(\ref{MI_0d})
and the leading exponents of the dynamic response functions.
\end{thebibliography}
\end{document}